\documentclass[aps, showpacs, showkeys,nofootinbib,floatfix]{revtex4}

\usepackage{amssymb}
\usepackage{amsmath}
\usepackage{graphicx}

\begin{document}

\title{Time Variable Cosmological Constant from Renormalization Group Equations}

\author{Lixin Xu}
\email{lxxu@dlut.edu.cn}

\affiliation{Institute of Theoretical Physics, School of Physics \&
Optoelectronic Technology, Dalian University of Technology, Dalian,
116024, P. R. China}

\begin{abstract}
In this paper, a time variable cosmological constant (CC) from
renormalization group equations (RGEs) is explored, where the
renormalization scale $\mu^2=R^{-2}_{CC}=Max(\dot{H}+2H^2,-\dot{H})$
is taken. The cosmological parameters, such as dimensionless energy
density, deceleration parameter and effective equation of state of
CC etc, are derived. Also, the cosmic observational constraints are
implemented to test the model's consistence. The results show that
it is compatible with cosmic data. So, it would be a viable dark
energy model.
\end{abstract}

\pacs{98.80.Es; 95.36.+x;98.80.-k}

\keywords{renormalization group equation; cosmological constant; dark energy}
\hfill TP-DUT/2009-06

\maketitle

\section{Introduction}

Cosmological constant (CC) is a long standing issue in cosmology and
physics. It was first introduced by Einstein to realize a static
universe about a century ago. However, it was found that this
space-time was unstable and not consistent with observed
cosmological expansion. Recently, CC returns to cosmology as a
natural candidate to dark energy to explain recent cosmic
observation that our universe is undergoing an accelerated expansion
firstly deduced from observational results of type Ia supernovae
\cite{ref:Riess98,ref:Perlmuter99}. In the context of quantum field
theory (QFT), CC has relations with the vacuum or zero point energy
density of quantum fields, via
\begin{equation}
\rho_{\Lambda}=\frac{1}{2}\int^{\Lambda}_{0}\frac{4\pi k^2
dk}{(2\pi)^3}\sqrt{k^2+m^2}\approx\frac{\Lambda^4}{16\pi^2},
\end{equation}
where $\Lambda\gg m$ is a UV cut-off. To balance an assumed UV
cut-off $\Lambda$ and the observational smallness of CC, tremendous
fine tuning is required. This is the so-called cosmological constant
problem \cite{ref:ccproblem}. As known, renormalization in QFT can
handle the infinities, and a dependence of the renormalized
constants on some energy scale $\mu$ is leaded. This renormalized
scale is usually identified with external momentum or characteristic
scale of the environment.

QFT in curved space-time leads to an infinite effective action or
vacuum expectation values (VEV) of the energy-momentum tensors of
the fields. A renormalization treatment can yield a scale-dependent
or running CC and a running Newton constant. Though, the absolute
values can not be calculated, the change with respect to the
renormalization scale can be calculated via RGEs originating in QFT
\cite{ref:RGEsQFT} and quantum gravity \cite{ref:RGEsQG}. In the
cosmological context, it is reasonable to identify the
renormalization scale with some characteristic scales of cosmology.
The related investigation can be found in \cite{ref:PhDthesis},
where the renormalization scale $\mu$ was given by the Hubble scale
$H$, the inverse radius $R^{-1}$ of the cosmological event horizon
and the inverse radius $T^{-1}$ of the particle horizon. A scaling
law having decoupling behavior at low energy in \cite{ref:PhDthesis}
(which was extensively studied in \cite{ref:RGEsQFT}) was explored
where the corresponding $\beta$-function for $\Lambda$ is
\begin{equation}
\mu\frac{d\Lambda}{d\mu}=A\mu^2,
\end{equation}
where $A\sim \pm M^2$ is determined by the masses $M$ and the spins
of the fields. Assuming constant masses and $\mu_0\sim H_0$, one has
\cite{ref:PhDthesis}
\begin{equation}
\frac{\Lambda(\mu)}{\Lambda_{0}}=L_0+L_1\frac{\mu^2}{\mu_0}, \quad
L_1\sim \pm\frac{M^2}{M^2_P},\label{eq:Lambdamu}
\end{equation}
where $L_0=1-L_1$. For sub-Planckian masses $M$, $|L_1|\ll 1$. Then
$L_0\sim 1$ as a model parameter can be teste by cosmic
observations. Obviously, when $L_0=1$, the CC becomes scale
independent and a real CC is recovered.

Another consideration of time variable CC can be found in
\cite{ref:Horvat1,ref:Horvat2,ref:Feng,ref:CCXU1,ref:CCXU2}.
Following the work \cite{ref:PhDthesis}, we are going to take
$R^{-2}_{CC}=Max(\dot{H}+2H^2,-\dot{H})$ as a possible candidate to
a renormalization scale to investigate the evolution of our universe
in this work when the Newton constant is fixed. Though, the reports
of \cite{ref:noRGE} have shown that the cosmology is not a RG flow
where RG was checked by a massless, minimally coupled scalar with a
quartic selfinteraction on a nondynamical, locally de Sitter
background. But it does not preclude using the RG conventionally to
relate quantities at different constant scales. So, $R^{-2}_{CC}$
can potentially be used as a renormalization scale. Here,
$R^{-2}_{CC}=Max(\dot{H}+2H^2,-\dot{H})$ is the causal connection
scale for spacially flat universe \cite{ref:CaiCC}. That was
investigated in the context of holographic dark energy
\cite{ref:CaiCC} and Ricci dark energy
\cite{ref:Gao,ref:HRDConstraint}. Furthermore, it was found that
only the case where $R^{-2}_{CC}=\dot{H}+2H^2$ as an IR cut-off was
consistent with the current cosmological observations when the
vacuum density appears as an independently conserved energy
component \cite{ref:CaiCC}. However, when a time variable CC is
considered these two cases must be checked over again for its
coupling with cold dark matter. It is just the case that will be
explored in this paper.

This paper is structured as follows. In Section \ref{sec:RGEsVCC},
we give a brief review of time variable CC. We will discuss the
evolution of time variable CC and derive the cosmological parameters
in Section \ref{sec:ECC}. Cosmic observational constraints are
implemented in Section \ref{sec:COC}. Section \ref{sec:con} are
conclusions.

\section{Time variable CC} \label{sec:RGEsVCC}

The Einstein equation with a cosmological constant is written as
\begin{equation}
R_{\mu\nu}-\frac{1}{2}Rg_{\mu\nu}+\Lambda g_{\mu\nu}=8\pi G
T_{\mu\nu},\label{eq:EE}
\end{equation}
where $T_{\mu\nu}$ is the energy-momentum tensor of ordinary matter
and radiation. From the Bianchi identity, one has the conservation
of the energy-momentum tensor $\nabla^{\mu}T_{\mu\nu}=0$, it follows
necessarily that $\Lambda$ is a constant. To have a time variable
cosmological constant $\Lambda=\Lambda(t)$, one can move the
cosmological constant to the right hand side of Eq. (\ref{eq:EE})
and take $\tilde{T}_{\mu\nu}=T_{\mu\nu}-\frac{\Lambda(t)}{8\pi
G}g_{\mu\nu}$ as the total energy-momentum tensor. Once again to
preserve the Bianchi identity or local energy-momentum conservation
law, $\nabla^{\mu}\tilde{T}_{\mu\nu}=0$, one has, in a spacially
flat FRW universe,
\begin{equation}
\dot{\rho}_{\Lambda}+\dot{\rho}_{m}+3H\left(1+w_{m}\right)\rho_{m}=0,\label{eq:conservation}
\end{equation}
where $\rho_{\Lambda}=M^2_{P}\Lambda(t)$ is the energy density of
time variable cosmological constant and its equation of state is
$w_{\Lambda}=-1$, and $w_{m}$ is the equation of state of ordinary
matter, for dark matter $w_m=0$. It is natural to consider
interactions between variable cosmological constant and dark matter
\cite{ref:Horvat2}, as seen from Eq. (\ref{eq:conservation}). After
introducing an interaction term $Q$, one has
\begin{eqnarray}
\dot{\rho}_{m}+3H\left(1+w_{m}\right)\rho_{m}=Q,\label{eq:rhom} \\
\dot{\rho}_{\Lambda}+3H\left(\rho_{\Lambda}+p_{\Lambda}\right)=-Q,\label{eq:rholambda}
\end{eqnarray}
and the total energy-momentum conservation equation
\begin{equation}
\dot{\rho}_{tot}+3H\left(\rho_{tot}+p_{tot}\right)=0.
\end{equation}
For a time variable cosmological constant, the equality
$\rho_{\Lambda}+p_{\Lambda}=0$ still holds. Immediately, one has the
interaction term $Q=-\dot{\rho}_{\Lambda}$ which is different from
the interactions between dark matter and dark energy considered in
the literatures \cite{ref:interaction} where a general interacting
form $Q=3b^2H\left(\rho_{m}+\rho_{\Lambda}\right)$ is put by hand.
With observation to Eq. (\ref{eq:rholambda}), the interaction term
$Q$ can be moved to the left hand side of the equation, and one has
the effective pressure of the time variable cosmological constant-
dark energy
\begin{equation}
\dot{\rho}_{\Lambda}+3H\left(\rho_{\Lambda}+p^{eff}_{\Lambda}\right)=0,
\end{equation}
where $p^{eff}_{\Lambda}=p_{\Lambda}+\frac{Q}{3H}$ is the effective
dark energy pressure. Also, one can define the effective equation of
state of dark energy, (for other definition, please see
\cite{ref:EEoS}),
\begin{eqnarray}
w^{eff}_{\Lambda}&=&\frac{p^{eff}_{\Lambda}}{\rho_{\Lambda}}\nonumber\\
&=&-1+\frac{Q}{3H\rho_{\Lambda}}\nonumber\\
&=&-1-\frac{1}{3}\frac{d \ln \rho_{\Lambda}}{d\ln a}.\label{eq:EEOS}
\end{eqnarray}
The Friedmann equation as usual can be written as, in a spacially
flat FRW universe,
\begin{equation}
H^2=\frac{1}{3M^2_P}\left(\rho_{m}+\rho_{\Lambda}\right)\label{eq:FE}.
\end{equation}

\section{Evolution of Time Variable CC and Cosmological
Parameters}\label{sec:ECC}

The time variable CC and $G$ was explored in \cite{ref:PhDthesis},
where the renormalization scale $\mu$ was given by the Hubble scale
$H$, the inverse radius $R^{-1}$ of the cosmological event horizon
and the inverse radius $T^{-1}$ of the particle horizon. In this
paper, we are going to reconsider the time variable CC when
$R^{-2}_{CC}=Max(\dot{H}+2H^2,-\dot{H})$ is taken as a
renormalization scale $\mu$, say $\mu^2= R^{-2}_{CC}$. Though the
authors of \cite{ref:CaiCC} have claimed the case where
$R^{-2}_{CC}=\dot{H}+2H^2$ as an IR cut-off was consistent with the
current cosmological observations when the vacuum density appears as
an independently conserved energy component. For the existence of
effective interaction between time variable CC and cold dark matter,
the two cases must been checked over again. We name the case
$\mu^2=\dot{H}+2H^2$ Model A and $\mu^2=-\dot{H}$ Model B.

\subsection{Molde A: $\mu^2=\dot{H}+2H^2$}\label{subsec:A}

In this case, the time variable CC can be written as
\begin{eqnarray}
\Lambda(t)&=&\Lambda_{0}(L_0 +L_1
\frac{R}{R_0})\nonumber\\
&=&\Lambda_{0}(L_0+L_1 \frac{\dot{H}+2H^2}{\dot{H}_{0}+2H^2_{0}}).
\end{eqnarray}
From the above equation, it seems the CC behaves quite like Ricci
dark energy. But in fact, it is different from that for its
effective interaction with cold dark matter. The corresponding
vacuum energy density is
\begin{eqnarray}
\rho_{\Lambda}&=&M^2_P\Lambda(t)\nonumber\\
&=&M^2_P\Lambda_{0}(L_0+L_1 \frac{\dot{H}+2H^2}{\dot{H}_{0}+2H^2_{0}})\nonumber\\
&=&M^2_P\Lambda_{0}\left[L_0+M_0
(\dot{H}+2H^2)\right],\label{eq:rholamH}
\end{eqnarray}
where $M_0=(1-L_0)/(\dot{H}_{0}+2H^2_{0})$. For its interaction
between $\rho_{\Lambda}$ and cold dark matter $\rho_{m}$, they are
not conservative separately. By using the definition of
dimensionless density parameters $\Omega_{m}(z)=\rho_{m}/(3M^2_P
H^2)$ and $\Omega_{\Lambda}(z)=\rho_{\Lambda}/(3M^2_P H^2)$, one has
\begin{eqnarray}
\rho_{m}&=&\left(\frac{1-\Omega_{\Lambda}}{\Omega_{\Lambda}}\right)\rho_{\Lambda}.
\end{eqnarray}
Then after easy algebra, the conservation Eq.
(\ref{eq:conservation}) can be rewritten as
\begin{equation}
\dot{H}+\frac{3}{2}H^2(1-\Omega_{\Lambda})=0.\label{eq:conHO}
\end{equation}
From Eq. (\ref{eq:rholamH}), one has the expression of
$\Omega_{\Lambda}$
\begin{eqnarray}
\Omega_{\Lambda}&=&\frac{\Lambda_{0}}{3H^2}\left[L_0+M_0
(\dot{H}+2H^2)\right]\nonumber\\
&=&\frac{B_0}{3H^2}+\frac{A_0}{3}\left(2+\frac{\dot{H}}{H^2}\right),\label{eq:omelaH}
\end{eqnarray}
where $A_0=\Lambda_0 M_0$ and $B_0=\Lambda_0 L_0$ for convenience.
Combining Eq. (\ref{eq:conHO}) and Eq. (\ref{eq:omelaH}), one has
\begin{equation}
\left(1-\frac{A_0}{2}\right)\frac{\dot{H}}{H^2}-\frac{B_0}{2}\frac{1}{H^2}+\frac{3}{2}\left(1-\frac{2A_0}{3}\right)=0,\label{eq:Ht}
\end{equation}
In terms of redshift $z$, the above Eq. (\ref{eq:Ht}) can be
rewritten as
\begin{equation}
-\left(1-\frac{A_0}{2}\right)\frac{(1+z)}{2}\frac{d\ln
E(z)}{dz}-\frac{B^{\prime}_0}{2}\frac{1}{E(z)}+\frac{3}{2}\left(1-\frac{2A_0}{3}\right)=0,\label{eq:Hz}
\end{equation}
where $H^2(z)=H^2_0E(z)$ is used and $B^{\prime}_0=B_0/H^2_0$ which
is a dimensionless parameter. The above differential equation has
the integration
\begin{equation}
E(z)=\left[1-\frac{B^{\prime}_0}{(3-2
A_0)}\right](1+z)^{3+A_0/(A_0-2)}+\frac{B^{\prime}_0}{(3-2
A_0)}.\label{eq:FDEQ1}
\end{equation}

\subsection{Model B: $\mu^2=-\dot{H}$}

In this case, the time variable CC can be written as
\begin{equation}
\Lambda(t)=\Lambda_{0}(L_0 +L_1 \frac{\dot{H}}{\dot{H}_0}).
\end{equation}
As done in \ref{subsec:A}, one has the expression of
$\Omega_{\Lambda}$ and differential equation of $H$
\begin{eqnarray}
\Omega_{\Lambda}=\frac{\Lambda_0}{3H^2}\left(L_0+N_0\dot{H}\right),\\
(1-\frac{C_0}{2})\frac{\dot{H}}{H^2}-\frac{B_0}{2}\frac{1}{H^2}+\frac{3}{2}=0,\label{eq:HdotH}
\end{eqnarray}
where $N_0=(1-L_0)/\dot{H}_0$ and $C_0=\Lambda_0 N_0$. Also, one can
find the Hubble parameter $H(z)$ as a solution of Eq.
(\ref{eq:HdotH}) with respect to redshift $z$ as follows
\begin{equation}
H^2(z)=H^2_0\left[\left(1-\frac{B^{\prime}_0}{3}\right)(1+z)^{\frac{6}{2-C_0}}+\frac{B^{\prime}_0}{3}\right].\label{eq:FDEQ2}
\end{equation}
It is clear that $\Lambda$CDM is recovered when $C_0=0$, i.e.
$L_0=1$.

\subsection{Discussion}

From these Friedmann equations (\ref{eq:FDEQ1}) and Eq.
(\ref{eq:FDEQ2}), one can immediately find out that the first terms
of the right hand of the equations behave like cold dark matter for
$A_0=0$ and $C_0=0$ respectively, i.e. $L_0=1$. In these cases, the
$\Lambda$CDM universe are recovered as expected in introduction.
These models contain two parameters $A_0$($C_0$) and $B^{\prime}_0$
which can be determined by cosmic observations. If this model does
not badly depart from $\Lambda$CDM universe, we can estimate the
values of parameters $A_0$($C_0$) and $B^{\prime}_0$. It is to say
$A_0\sim 0$($C_0\sim 0$) and $B^{\prime}_0\sim
3\Omega_{\Lambda0}\sim 2.1$ which can be tested by cosmic
observations. In terms of redshift, the deceleration parameter and
effective EoS of CC can be written as
\begin{eqnarray}
q&=&-1+\frac{(1+z)}{2}\frac{d\ln H^2}{dz},\label{eq:qz}\\
w^{eff}_{\Lambda}&=&-1+\frac{(1+z)}{3}\left(\frac{d\ln
H^2}{dz}+\frac{d\ln \Omega_{\Lambda}(z)}{dz}\right).\label{eq:wz}
\end{eqnarray}

From Eq. (\ref{eq:FDEQ1}), one can find two singularity points with
parameter values of $A_0=2\text{ and } 3/2$ respectively. The same
case can be found in Eq. (\ref{eq:FDEQ2}) when $C_0=2$. When
$A_0=2$, the first term of left hand side of Eq. (\ref{eq:Ht})
vanishes. Then one has a constant Hubble parameter $H^2=-1/B_0$. For
the positivity of the value of $B_0>0$, it does not describe a
physical system. However, for $C_0=2$, one has $H^2=B_0/3$. Then, in
this case, a de Sitter or anti de Sitter universe can be obtained.
When $A_0=3/2$, one has $\dot{H}=2B_0$ which corresponds to scale
factor $a(t)\sim \exp(B_0t^2)$.

\section{Cosmic Observational Constraints}\label{sec:COC}

\begin{figure}[tbh]
\centering
\includegraphics[width=5.0in]{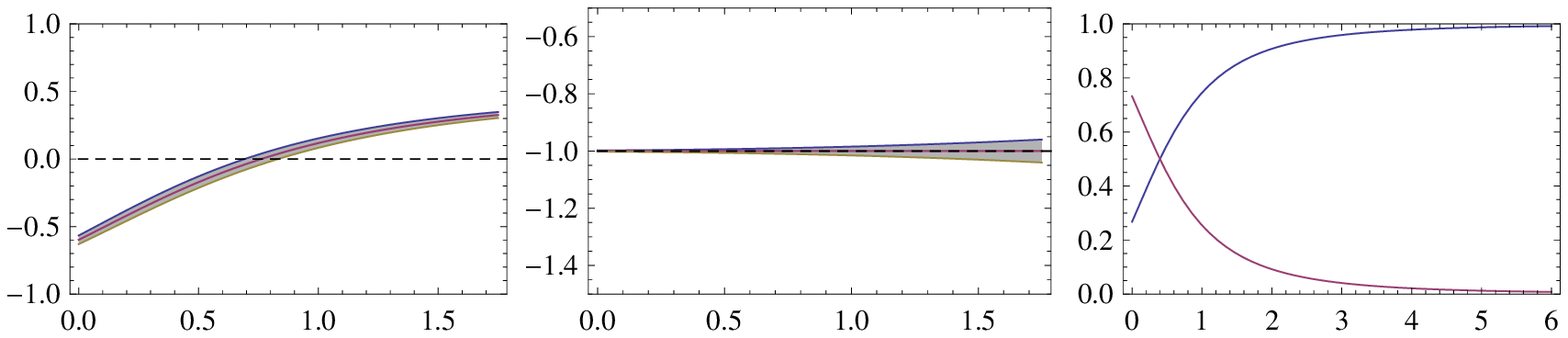}
\caption{Model A: The evolution curves of $q(z)$ (left panel),
$w^{eff}_{\Lambda}(z)$ (central panel) with $1\sigma$ error region
and dimensionless parameters $\Omega_{m}(z)$ and
$\Omega_{\Lambda}(z)$ (right panel) with respect to redshift $z$
where the best fit values are adopted.}\label{fig:qwomegaA}
\end{figure}

\begin{figure}[tbh]
\centering
\includegraphics[width=5.0in]{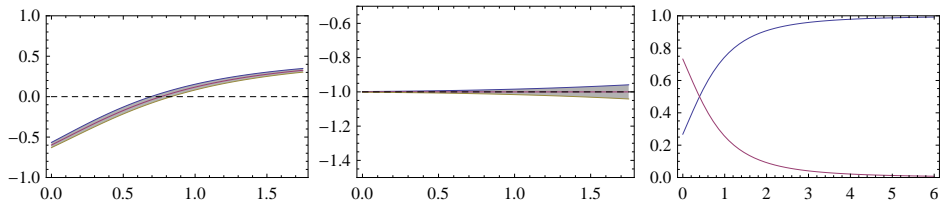}
\caption{Model B: The evolution curves of $q(z)$ (left panel),
$w^{eff}_{\Lambda}(z)$ (central panel) with $1\sigma$ error region
and dimensionless parameters $\Omega_{m}(z)$ and
$\Omega_{\Lambda}(z)$ (right panel) with respect to redshift $z$
where the best fit values are adopted.}\label{fig:qwomegaB}
\end{figure}

\begin{figure}[tbh]
\centering
\includegraphics[width=2.0in]{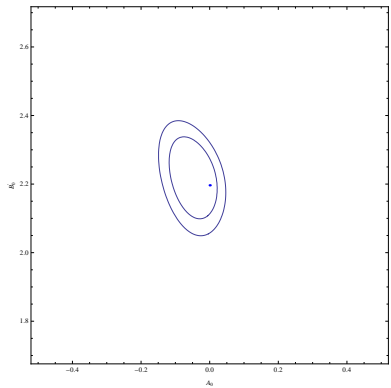}
\caption{Model A: The contours in the planes of $A_0-B^{\prime}_0$
with $1\sigma$ and $2\sigma$ regions. The central dot denotes the
best fit values of model parameters.}\label{fig:consA}
\end{figure}

\begin{figure}[tbh]
\centering
\includegraphics[width=2.0in]{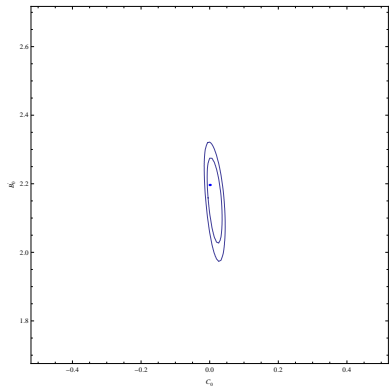}
\caption{Model B: The contours in the planes of $C_0-B^{\prime}_0$
with $1\sigma$ and $2\sigma$ regions. The central dot denotes the
best fit values of model parameters.}\label{fig:consB}
\end{figure}

Now, it is proper to present the constraint results by using cosmic
observations: SN Ia, BAO and CMB shift parameter $R$, for the
details please see Appendix \ref{sec:append}. In this work, 397 SN
Ia Constitution dataset, the ratio $D_{V}(0.35)/D_{V}(0.2)$ detected
by BAO and CMB $R$ from WMAP5 are used. After the calculation as
described in Appendix \ref{sec:append}, the results are listed in
Tab. \ref{tab:result}.
\begin{table}[tbh]
\begin{center}
\begin{tabular}{c|c|c|c|c|c|c}
\hline\hline
Model & $\chi^2_{min}$ & $\Omega_{\Lambda0}(1\sigma)$ & $A_0 \text{ or } C_0 (1\sigma)$ & $B^{\prime}_0(1\sigma)$ & $z_T(1\sigma)$ &$\chi^2_{min}/dof$\\
\hline A & $473.593$  & $0.732\pm 0.021$  & $(1.0\times 10^{-11})^{+0.0217}_{-0.108}$ & $2.196^{+0.0780}_{-0.0833}$ & $0.762^{+0.071}_{-0.069}$ & $1.193$ \\
\hline B & $473.593$  & $0.732\pm 0.021$  & $(-7.011\times 10^{-10})^{+0.0292}_{-0.00703}$ & $2.196^{+0.0780}_{-0.0833}$ & $0.762^{+0.072}_{-0.070}$ & $1.193$ \\
\hline\hline
\end{tabular}
\caption{The minimum values of $\chi^2$ and best fit values of the
parameters. Here $z_T$ is the transition redshift from decelerated
expansion to accelerated expansion and $dof$ denotes the model
degrees of freedom.}\label{tab:result}
\end{center}
\end{table}
The evolution curves of $q(z)$, $w^{eff}_{\Lambda}(z)$ and
dimensionless density parameters $\Omega_{m}(z)$ and
$\Omega_{\Lambda}(z)$ are plotted in Fig. \ref{fig:qwomegaA} and
Fig. \ref{fig:qwomegaB}.
Also, the contour plots of $A_{0}-B^{\prime}_0$ and
$C_0-B^{\prime}_0$ are shown in Fig. \ref{fig:consA} and Fig.
\ref{fig:consB}.
From the left panels of Fig. \ref{fig:qwomegaA} and Fig.
\ref{fig:qwomegaB}, one can easily find that our universe is
undergoing accelerated expansion at late time, and the transition
redshift from decelerated expansion to accelerated expansion are
$z_T\sim 0.762$, which are consistent with other analysis results
with best fit parameter values. In the early epoch, our universe is
dominated by cold dark matter, that can be seen from the left and
right panels of Fig. \ref{fig:qwomegaA} and Fig. \ref{fig:qwomegaB}
where the deceleration parameter is $q\rightarrow 1/2$ (dark matter
dominated) and $\Omega_{m}\rightarrow 1$ at high redshift. From the
central panels of Fig. \ref{fig:qwomegaA} and Fig.
\ref{fig:qwomegaB}, one can see the effective EoS of time variable
CC is almost constant $w^{eff}_{\Lambda}(z)\sim -1$. So, the
universe is quasi-$\Lambda$CDM and it is not necessary to worry
about the structure formation of the universe. With the best fit
values of the parameters and
$L_0=B^{\prime}_0/(3\Omega_{\Lambda0})$, one obtains the mass of the
fields is $M=|1-L_0|^{1/2} M_P\sim 1.490\times 10^{-8} M_P$ and
$L_1\sim -2.220\times 10^{-16}$ in Model A and Model B which
confirms the prediction in the introduction.

\section{Conclusions}\label{sec:con}

In this paper, time variable cosmological constant from
renormalization group equations (RGEs) is explored, where the
renormalization scale $\mu^2=R^{-2}_{CC}=Max(\dot{H}+2H^2,-\dot{H})$
is taken. The cosmological parameters, such as dimensionless energy
density, deceleration parameter and effective EoS of CC etc, are
derived. Also, the comic observational constraints are implemented
to test the model's consistence, the results are shown in Tab.
\ref{tab:result}. As investigated, with this time variable CC, the
universe is undergoing an accelerated expansion at late time and a
decelerated expansion at high redshift. And, the transition redshift
from decelerated expansion to accelerated expansion is $z_T\sim
0.762$ which is consistent with other results. The effective EoS of
time variable CC is almost constant $w^{eff}_{\Lambda}(z)\sim -1$.
In early epoch, our universe is dominated by cold dark matter, that
can be seen from the left and right panels of Fig.
\ref{fig:qwomegaA} and Fig. \ref{fig:qwomegaB} where the
deceleration parameter is $q\rightarrow 1/2$ and
$\Omega_{m}\rightarrow 1$ (dark matter dominated) at high redshift.
And the current cold dark matter density ratio is
$\Omega_{\Lambda0}\sim 0.732$ which is also compatible with other
analysis. So, via RGEs with the renormalization scale
$\mu^2=R^{-2}_{CC}=Max(\dot{H}+2H^2,-\dot{H})$, time variable CC is
viable. With the best fit values of the parameters and
$L_0=B^{\prime}_0/(3\Omega_{\Lambda0})$, one obtains the mass of the
fields is $M=|1-L_0|^{1/2} M_P\sim 1.490\times 10^{-8} M_P$ and
$L_1\sim -2.220\times 10^{-16}$ in Model A and Model B which
confirms the prediction in the introduction and implies that the
mass scale having relations with CC about $10^{-8}$ order of $M_P$.

\acknowledgements{This work is supported by NSF (10703001), SRFDP
(20070141034) of P.R. China.}

\appendix

\section{Cosmic Observations}\label{sec:append}

\subsection{SN Ia}
We constrain the parameters with the 397 SN Ia Constitution dataset
including $397$ SN Ia \cite{ref:Condata}. Constraints from SN Ia can
be obtained by fitting the distance modulus $\mu(z)$
\begin{equation}
\mu_{th}(z)=5\log_{10}(D_{L}(z))+\mu_{0},
\end{equation}
where, $D_{L}(z)$ is the Hubble free luminosity distance $H_0
d_L(z)/c$ and
\begin{eqnarray}
d_L(z)&=&c(1+z)\int_{0}^{z}\frac{dz^{\prime}}{H(z^{\prime})}\\
\mu_0&\equiv&42.38-5\log_{10}h,
\end{eqnarray}
where $H_0$ is the Hubble constant which is denoted in a
re-normalized quantity $h$ defined as $H_0 =100 h~{\rm km ~s}^{-1}
{\rm Mpc}^{-1}$. The observed distance moduli $\mu_{obs}(z_i)$ of SN
Ia at $z_i$ is
\begin{equation}
\mu_{obs}(z_i) = m_{obs}(z_i)-M,
\end{equation}
where $M$ is their absolute magnitudes.

For SN Ia dataset, the best fit values of parameters in a model can
be determined by the likelihood analysis is based on the calculation
of
\begin{eqnarray}
\chi^2(p_s,m_0)&\equiv& \sum_{SNIa}\frac{\left[
\mu_{obs}(z_i)-\mu_{th}(p_s,z_i)\right]^2} {\sigma_i^2} \nonumber\\
&=&\sum_{SNIa}\frac{\left[ 5 \log_{10}(D_L(p_s,z_i)) - m_{obs}(z_i)
+ m_0 \right]^2} {\sigma_i^2}, \label{chi2}
\end{eqnarray}
where $m_0\equiv\mu_0+M$ is a nuisance parameter (containing the
absolute magnitude and $H_0$) that we analytically marginalize over
\cite{ref:SNchi2},
\begin{equation}
\tilde{\chi}^2(p_s) = -2 \ln \int_{-\infty}^{+\infty}\exp \left[
-\frac{1}{2} \chi^2(p_s,m_0) \right] dm_0 \; ,
\label{chi2_marginalization}
\end{equation}
to obtain
\begin{equation}
\tilde{\chi}^2 =  A - \frac{B^2}{C} + \ln \left(
\frac{C}{2\pi}\right) , \label{chi2_marginalized}
\end{equation}
where
\begin{equation}
A=\sum_{SNIa} \frac {\left[5\log_{10}
(D_L(p_s,z_i))-m_{obs}(z_i)\right]^2}{\sigma_i^2},
\end{equation}
\begin{equation}
B=\sum_{SNIa} \frac {5
\log_{10}(D_L(p_s,z_i)-m_{obs}(z_i)}{\sigma_i^2},
\end{equation}
\begin{equation}
C=\sum_{SNIa} \frac {1}{\sigma_i^2} \; .
\end{equation}
The Eq. (\ref{chi2}) has a minimum at the nuisance parameter value
$m_0=B/C$. Sometimes, the expression
\begin{equation}
\chi^2_{SNIa}(p_s,B/C)=A-(B^2/C)\label{eq:chi2SNIa}
\end{equation}
is used instead of Eq. (\ref{chi2_marginalized}) to perform the
likelihood analysis. They are equivalent, when the prior for $m_0$
is flat, as is implied in (\ref{chi2_marginalization}), and the
errors $\sigma_i$ are model independent, what also is the case here.

To determine the best fit parameters for each model, we minimize
$\chi^2(p_s,B/C)$ which is equivalent to maximizing the likelihood
\begin{equation}
{\cal{L}}(p_s) \propto e^{-\chi^2(p_s,B/C)/2} .
\end{equation}

\subsection{BAO}
The BAO are detected in the clustering of the combined 2dFGRS and
SDSS main galaxy samples, and measure the distance-redshift relation
at $z = 0.2$. BAO in the clustering of the SDSS luminous red
galaxies measure the distance-redshift relation at $z = 0.35$. The
observed scale of the BAO calculated from these samples and from the
combined sample are jointly analyzed using estimates of the
correlated errors, to constrain the form of the distance measure
$D_V(z)$ \cite{ref:Okumura2007,ref:Eisenstein05,ref:Percival}
\begin{equation}
D_V(z)=\left[(1+z)^2 D^2_A(z) \frac{cz}{H(z)}\right]^{1/3},
\label{eq:DV}
\end{equation}
where $D_A(z)$ is the proper (not comoving) angular diameter
distance which has the following relation with $d_{L}(z)$
\begin{equation}
D_A(z)=\frac{d_{L}(z)}{(1+z)^2}.
\end{equation}
Matching the BAO to have the same measured scale at all redshifts
then gives \cite{ref:Percival}
\begin{equation}
D_{V}(0.35)/D_{V}(0.2)=1.812\pm0.060.
\end{equation}
Then, the $\chi^2_{BAO}(p_s)$ is given as
\begin{equation}
\chi^2_{BAO}(p_s)=\frac{\left[D_{V}(0.35)/D_{V}(0.2)-1.812\right]^2}{0.060^2}\label{eq:chi2BAO}.
\end{equation}

\subsection{CMB shift Parameter R}

The CMB shift parameter $R$ is given by \cite{ref:Bond1997}
\begin{equation}
R(z_{\ast})=\sqrt{\Omega_m H^2_0}(1+z_{\ast})D_A(z_{\ast})/c
\end{equation}
which is related to the second distance ratio
$D_A(z_\ast)H(z_\ast)/c$ by a factor $\sqrt{1+z_{\ast}}$. Here the
redshift $z_{\ast}$ (the decoupling epoch of photons) is obtained by
using the fitting function \cite{Hu:1995uz}
\begin{equation}
z_{\ast}=1048\left[1+0.00124(\Omega_bh^2)^{-0.738}\right]\left[1+g_1(\Omega_m
h^2)^{g_2}\right],
\end{equation}
where the functions $g_1$ and $g_2$ are given as
\begin{eqnarray}
g_1&=&0.0783(\Omega_bh^2)^{-0.238}\left(1+ 39.5(\Omega_bh^2)^{0.763}\right)^{-1}, \\
g_2&=&0.560\left(1+ 21.1(\Omega_bh^2)^{1.81}\right)^{-1}.
\end{eqnarray}
The 5-year {\it WMAP} data of $R(z_{\ast})=1.710\pm0.019$
\cite{ref:Komatsu2008} will be used as constraint from CMB, then the
$\chi^2_{CMB}(p_s)$ is given as
\begin{equation}
\chi^2_{CMB}(p_s)=\frac{(R(z_{\ast})-1.710)^2}{0.019^2}\label{eq:chi2CMB}.
\end{equation}

\end{document}